\documentclass[aps,pre,preprint,superscriptaddress,longbibliography]{revtex4-1}

\usepackage{natbib}
\usepackage{amsmath,amssymb,pifont,wasysym,MnSymbol,}
\usepackage{graphicx,color,subfigure,bm}
\usepackage{lastpage,fancyhdr}

\def\etal{\emph{et al. }}
\begin{document}

\title{Viscoelastic propulsion of a rotating dumbbell}

\def\r{{\bf r}}
\def\u{{\bf u}}
\def\De{{\rm De}}
\def\e{{\bf e}}
\def\U{{\bf U}}
\def\O{{\boldsymbol \Omega}}
\def\g{{\boldsymbol \gamma}}

\author{J. Amadeus Puente-Vel\'azquez}
\affiliation{Instituto de Investigaciones en Materiales, Universidad
Nacional Aut\'onoma de M\'exico, Apdo. Postal 70-360, Ciudad de M\'exico
04510, M\'exico}

\author{Francisco A. God\'inez}
\affiliation{Instituto de Ingenier\'ia, Universidad Nacional Aut\'onoma de
M\'exico, UNAM, C.U., 04510 Mexico City, Mexico}
\affiliation{Polo Universitario de Tecnolog\'ia Avanzada, niversidad Nacional Aut\'onoma de M\'exico, Apodaca 66629,
Nuevo Leon, Mexico}

\author{Eric Lauga}
\email{e.lauga@damtp.cam.ac.uk}
\affiliation{Department of Applied Mathematics and Theoretical Physics, Wilberforce Road, University of Cambridge, Cambridge CB3 0WA, UK}

\author{R. Zenit}
\email{zenit@unam.mx; zenit@brown.edu}
\affiliation{Instituto de Investigaciones en Materiales, Universidad Nacional Aut\'onoma de M\'exico, Apdo. Postal 70-360, Coyoacan, Ciudad de Mexico 04510, Mexico}
\affiliation{School of Engineering, Brown University, 184 Hope St, Providence, RI 02912, USA}

\begin{abstract}
Viscoelastic fluids impact the locomotion of swimming microorganisms and can be harnessed to devise new types of self-propelling devices. Here we  report on experiments demonstrating the use of normal stress differences for propulsion.
Rigid  dumbbells are rotated by an external magnetic field along their axis of symmetry in a Boger fluid. When the dumbbell is asymmetric  (snowman geometry), non-Newtonian  normal stress differences  lead to net propulsion in the direction of the smaller sphere. The use of a simple model allows to rationalise the experimental results and to predict the dependence of the snowman swimming speed on the size ratio between the two spheres.
\end{abstract}

\maketitle

\section{Introduction}

As a difference with well-studied Newtonian fluids, the rheological behaviour of complex fluids is often non-Newtonian and, with stresses  no longer proportional to the instantaneous value of the shear rate,  remarkable new physics emerge~\cite{Morrisonbook}.  Well-studied examples of complex fluids include polymers solutions~\cite{doi88}, polymer melts~\cite{larson88}, colloidal suspensions~\cite{gast1998simple}, gels~\cite{larson99} and liquid crystals~\cite{prost1995physics}. The mathematical modelling of the behaviour of complex fluids involves the combination of kinetic theory describing the microscopic thermodynamics of individual constituents in the solvent ~\cite{birdvol2,larson99} along  with the new fluid mechanics which results from it~\cite{bird76,birdvol1,tanner88,bird95}.

After many years of  research on the physics of complex fluids,
recent  efforts have focused on their impact in biological systems
and in particular on the locomotion of small organisms~\cite{spagnolie2015complex}.  For example, spermatozoa have to swim in cervical mucus, a complex fluid with a relaxation time on the order of seconds~\cite{katz80}, much larger than the typical oscillation time scale of a spermatozoon and its flagellum~\cite{gaffney11}. As a result, the flagellar waveforms of spermatozoa are  impacted by non-Newtonian stresses and they differ significantly from those observed in Newtonian fluids~\cite{ishijima1986flagellar}.
Theoretical efforts in the context of elastic fluids have allowed to compute leading-order hydrodynamic forces  on beating flagella~\cite{fulford1998}, their impact on the beating waveforms~\cite{Fu08} and on cell locomotion~\cite{lauga07,Fu09,teran2010}. Later work showed that the coupling between the flexibility of the swimmer and the stresses in the fluid could lead to both hindered and enhanced locomotion~\cite{riley14,thomases14}.
Collective effects are also affected and interactions of spermatozoa in mucus leads to long-lived clusters~\cite{tung17}.
Even in the absence of elastic stresses, shear-thinning rheology impacts waving locomotion~\cite{velez2013waving}.

Although  spermatozoa have received the biggest share of research, other small swimming organisms have been shown to be affected by complex fluids. Nematodes~\cite{arratia2011,gagnon13}, algae~\cite{qin15} and bacteria~\cite{patteson2015running,Liu2011,spagnolie13} exhibit a different behaviour  when  swimming in polymeric fluids. Some changes in the swimming motion can be in part explained by the ability of the cells to push against the microstructure~\cite{BergTurner1979,magariyama02,leshansky09} and by the presence of  length scales in the fluid comparable to, or larger than, the cell size~\cite{martinez14}. A boost in locomotion is also possible due to polymer depletion near the swimmer and resulting apparent fluid
slip~\cite{man2015phase,zottl2017enhanced}.  A commonly used model for microorganism locomotion is the spherical squirmer~\cite{Zhu2011} whose swimming has been shown to be
significantly affected by viscoelasticity~\cite{Zhu2012,Decorato2015}.  More broadly, suspensions of swimmers behave as complex fluids themselves as shown experimentally for bacteria~\cite{sokolov09,ryan11}
and algae~\cite{rafai2010effective,mussler2013effective}, and rationalised theoretically~\cite{hatwalne04,saintillan2018rheology}.

Beyond their impact on the  biological world,  complex fluids also affect artificial swimmers. Flexible filaments oscillated externally using  magnetic fluids  can swim  faster in  Boger fluids~\cite{laugazenit13} in a manner which depends   on the kinematics of their  deformation~\cite{godinez15}.  More importantly, the nonlinearities offered by  non-Newtonian flows can be exploited to induce novel propulsion modes and design new types of swimmers~\cite{lauga14}.  Rigid helices rotated externally and unable to move in water can swim efficiently in a gel~\cite{schamel14}. The time-reversibility constraints of Stoke flows can be bypassed and shear-dependent viscosity used to generate net forces from time-reversible forcing~\cite{qiu2014swimming}. Propulsive forces can also result  from the secondary flows induced by non-Newtonian normal stress differences~\cite{normand08,pak10,pak12,keim12}. Secondary flows have been long recognized as an important feature of viscoelatic flows\cite{birdvol1}; however, they have only been studied for a few canonical cases.

In this work we  report on experiments demonstrating the use of normal stress differences for propulsion following the theoretical proposal of
Ref.~\cite{pak12}. We use rigid  dumbbells  rotated by an external magnetic field along their axis of symmetry in a Boger fluid. By symmetry, no motion is induced for symmetric dumbbells but when  the shape of the dumbbell is asymmetric  (snowman geometry), non-Newtonian  normal stress differences lead to net propulsion. The use of a simple model allows to rationalise the experimental results and to predict the dependence of the snowman swimming speed as a function of the size ratio between the two spheres~\cite{pak12}. Note that a similar configuration was studied in Ref.~\cite{Rogowski2018}; however, in their case, the dumbbell  was composed of two equal-sized spheres and the propulsion resulted from heterogeneities of the media.

The paper is organised as follows. In
 \S\ref{sec:setup} we introduce the experimental setup, the  dumbbells used for the locomotion, and characterise the rheological properties of the fluids used. We   present our experimental results in \S\ref{sec:results} and compared them with the model of Ref.~\cite{pak12} in
\S\ref{sec:model}. A conclusion and perspectives are offered in
\S\ref{sec:conclusion}.

\section{Experimental setup and test fluids}
\label{sec:setup}

\begin{figure}
\centering
\includegraphics[width=0.4\columnwidth]{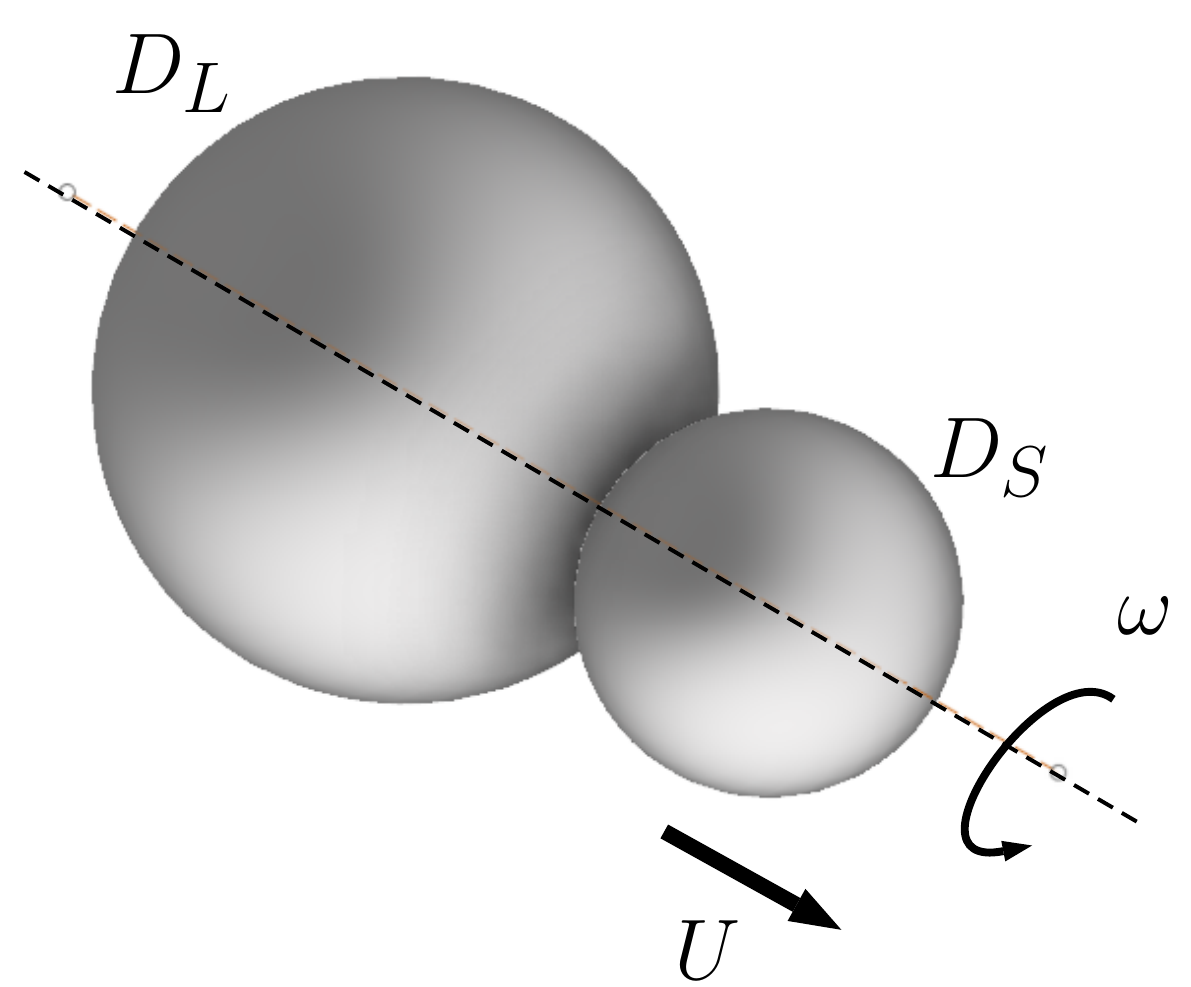}
\caption{Schematic view of the configuration studied experimentally.  A neutrally-buoyant dumbbell made up of two attached spherical particles of diameters $D_L$ and $D_S\leq D_L $ is  rotated with angular velocity $\omega$ along its symmetry axis in  a viscoelastic fluid. }
\label{fig:swimmer}
\end{figure}

Following the theoretical proposal put forward by Pak \etal~\cite{pak12}, we conduct  experiments consisting in rotating a neutrally-buoyant sphere-sphere magnetic dumbbell immersed in a viscoelastic fluid. The  experiment is schematically  represented in Fig.~\ref{fig:swimmer}. The large sphere in the dumbbell   has diameter $D_L$ and the smaller one $D_S$ and the whole dumbbell   rotates with angular velocity $\omega$ along its symmetry axis.

We use the magnetic setup developed in Ref.~\cite{godinez12}, which is a device  capable of producing a magnetic field of uniform strength 6 mT  and   rotated mechanically. The dumbbells were placed inside a rectangular tank (160~mm $\times$ 100~mm $\times$ 100~mm) that fit into the region of uniform magnetic field inside the coils of approximately $(100$~mm)$^3$ in size where the test fluids were contained.  In this configuration, the ratio of the large sphere to the containing walls was 0.09; the wall effects are therefore small but finite.

The spheres in the dumbbells were made out of plastic and we inserted inside them  a permanent magnet   (Magcraft, models NSN0658). In  all cases, the angular frequency of the rotating coils was below the step-out frequency~\cite{godinez12} and the dumbbells rotate at the same rate as the external magnetic field.  The swimmers were nearly neutrally buoyant; the effective density was adjusted by filling up the cavity (where the permanent magnet was inserted) with small amounts of the test liquid.

In order to achieve locomotion in the viscoelastic fluid, the sphere-sphere arrangement had to be made of  spheres with unequal sizes (snowman configuration). Five dumbbells were considered, varying the diameter ratio $D^*=D_S/D_L$ from 0 to 1. The dimensions  for each dumbbell are shown in Table \ref{sphereproperties}.  The motion of the dumbbell was filmed and its position was tracked in time.

\begin{table}[b]
  \centering
  \begin{tabular}{ c | c | c | c }
  \hline\hline
\hspace{3mm}  {\bf Swimmer}\hspace{3mm}  &\hspace{3mm} $D^*=D_S/D_L$ \hspace{3mm} & \hspace{3mm}diameter, $D_L$ \hspace{3mm}&\hspace{3mm} diameter, $D_S\hspace{3mm}$ \\
      (N,B) & (-) &(mm) & (mm)   \\ \hline
      ({\color{red}$\circ$, $\bullet$})& 0 & 9.0 & 0.0 \\
      ({\color{blue}$\lozenge$,$\blacklozenge$})& 0.27 & 7.7 & 2.1  \\
     ({\color{magenta}$\triangledown$,$\blacktriangledown$})& 0.64& 5.9 & 3.8 \\
      ({\color{green}$\triangleleft$,$\blacktriangleleft$})& 0.82& 9.5 &  8.0\\
 ($\Box$, $\blacksquare$)& 1 & 9.0 & 9.0 \\
       \hline\hline
  \end{tabular}
  \caption{Sizes of the spheres in the five dumbbells used in this investigation.}
  \label{sphereproperties}
\end{table}

\begin{figure}
\centering
\includegraphics[width=0.6\textwidth]{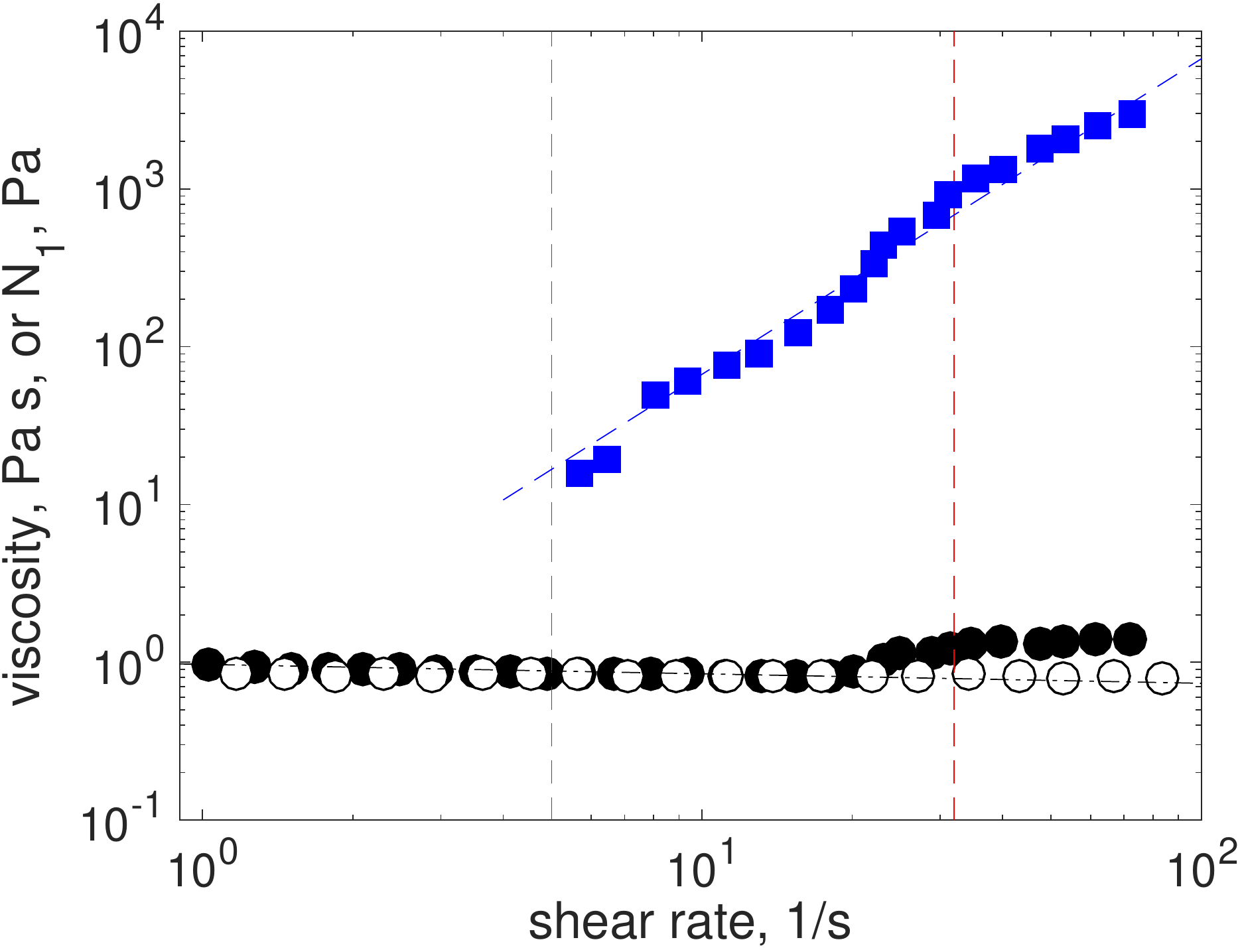}
\caption{Circles: Shear viscosity (in Pa.s) as a function of shear rate (1/s) for the Boger (B, $\bullet$) and Newtonian (N, $\circ$) liquids. The dashed dotted line shows a fit to a power-law viscosity fit with $m=0.97$ Pa s$^n$ and $n=0.94$,  (Eq.~\ref{eqn:PowerLaw}) for $\dot{\gamma}<20$ s$^{-1}$. Squares: First normal stress difference, $N_1$, ({\color{blue} $\blacksquare$}), as a function of shear rate. The dashed blue line shows the fit to an Oldroyd-B model  (Eq.~\ref{eqn:OlroydB}). The two vertical lines show the window of shear rates within which the experiments were conducted ($5.5<\omega<31.0$ s$^{-1}$).}
\label{fig:rheology}
\end{figure}

Two types of fluids were fabricated, tested and used. One was Newtonian (N)  and used as a reference fluid while the other was a Boger-type fluid (B)   with a nearly constant shear viscosity and  viscoelastic properties~\cite{boger1977}. The B fluid was fabricated by dissolving polyacrylamide (PAA, molecular weight 5$\times 10 ^{6}$ g/mol) in non-ionic water at a slow mixing rate for 24 hours. Afterwards, the solution was added to glucose and slowly mixed for four days. The proportions (by weight) of the mixture were 84.9\% glucose, 15\% water and 0.1\% PAA. The liquid was left in repose for two weeks previous to testing.

The rheological properties were determined in an Anton Paar rheometer (stress controlled and cone-plate geometry).   All physical and rheological properties of both fluids are summarised in Table~\ref{rheologicalproperties}.
Steady shear tests were conducted, within the range of shear rates within $ 1 < \dot\gamma < 100$ s$^{-1}$, to determine the shear viscosity and the first normal stress difference. All tests were conducted at $T=23^\circ$C. The results of the rheological tests are shown in Fig.~\ref{fig:rheology} (filled symbols). After the shear viscosity of the B fluid was determined, the N fluid was prepared by adding water to glucose until the viscosity was matched to that of the B fluid. The N fluid was left standing for 24 hours prior to testing. The shear viscosity of the N fluid is also shown in Fig.~\ref{fig:rheology} (empty circles).

The viscosity measurements show that the B fluid has a nearly constant viscosity, $\eta$, for shear rates $\dot\gamma < 30$ s$^{-1}$. Fitting the data to a power-law model as
\begin{equation}\label{eqn:PowerLaw}
  \eta=m {\dot\gamma}^{n-1},
\end{equation}
leads to $m=0.97$ Pa s$^n$ and $n=0.94$. Hence,  shear-tinning effects are almost negligible. At higher shear rates, the fluid displayed a slight shear-thickening behaviour. The range of shear rates attainable during the experiments (shown by the two vertical lines in the figure) are within the nearly-constant viscosity region.

\begin{table}[b]
  \centering
  \begin{tabular}{ c | c | c | c | c | c }\hline\hline
    \hspace{2mm} {\bf Fluid}     \hspace{2mm}  & $\rho$ & $\eta_o$ &    \hspace{2mm}  $n$    \hspace{2mm}  & $    \hspace{2mm} \lambda$     \hspace{2mm} &     \hspace{2mm} $\beta$    \hspace{2mm}  \\
     &     \hspace{2mm} kg/m$^3$    \hspace{2mm}   &    \hspace{2mm}  Pa s     \hspace{2mm} & (-) & s & (-)\\ \hline
       N  & 1510 & 0.84 & 1 & - & - \\
       B  &     \hspace{2mm} 1508    \hspace{2mm}  &     \hspace{2mm} 0.84     \hspace{2mm} &     \hspace{2mm} 0.94    \hspace{2mm}  &     \hspace{2mm} 0.5102     \hspace{2mm} &     \hspace{2mm} 0.1149    \hspace{2mm}  \\
       \hline\hline
  \end{tabular}
  \caption{Physical and rheological properties of the fluids used in this investigation.}
  \label{rheologicalproperties}
\end{table}

To characterise the viscoelastic nature of the B fluid, we fitted the measurement of the first normal stress difference,  $N_1$, to an Oldroyd-B model as  \cite{oldroyd1950}
\begin{equation}\label{eqn:OlroydB}
  N_1=2\eta_o (1-\beta) \lambda {\dot\gamma}^2,
\end{equation}
where $\eta_o=\eta_p+\eta_s$ is the total viscosity (where $\eta_p$ and $\eta_s$ are the polymer and solvent viscosities, respectively), $\beta=\eta_s/\eta_o$ is the ratio of solvent to total viscosities and $\lambda$ is the relaxation time. Since the rheological measurements agreed well with the ${\dot\gamma}^2$-dependence of $N_1$, we estimate  the relaxation time, $\lambda$, from fitting the data to Eq.~(\ref{eqn:OlroydB}). We obtain $\lambda\approx 0.5102$ s  with $\eta_o\approx 0.84$ Pa s, $\eta_s\approx 0.19$ Pa s and $\beta\approx 0.2252$. The resulting fit is  shown in Fig.\ref{fig:rheology}. With the determination of the relaxation we can define the Deborah number $\De$ as
\begin{equation}\label{eqn:De}
  \De=\omega \lambda,
\end{equation}
where $\omega$ is the rotational speed of the swimmer. For the experiments reported here the maximum $\De$ number was 15. The Reynolds number, defined as
\begin{equation}
  {\rm Re}=\frac{UD_L\rho}{\eta_o},
\end{equation}
had a maximum value of 0.11. For this range of values of $\rm Re$ and $\De$ we can assert that the elastic effects dominate, $\De/{\rm Re}>5$. Note that, since the flow is induced by rotation, a rotational Reynolds number can also be considered
\begin{equation}
  {\rm Re_\omega}=\frac{\omega D_L^2\rho}{\eta_o}.
\end{equation}
For the range of parameters tested in this study, $\rm Re_\omega<4.5$, for which the inertial forces are no longer negligible. However, the ratio $\De/{\rm Re_\omega}=3.5$ (independent of $\omega$) which indicates that the elastic effects remain dominant. For the case of particles in a viscoelastic fluid, flowing a tube, work in Refs.~\cite{Trofa2015,DelGiudice2015} has shown that for $\De/{\rm Re}>1$ the flow is dominated by elastic effects even   with finite inertia. Since this is the limit we are in, the  results we obtain are primarily due to fluid elasticity.

\begin{figure}[t]
\centering
\includegraphics[width=0.6\textwidth]{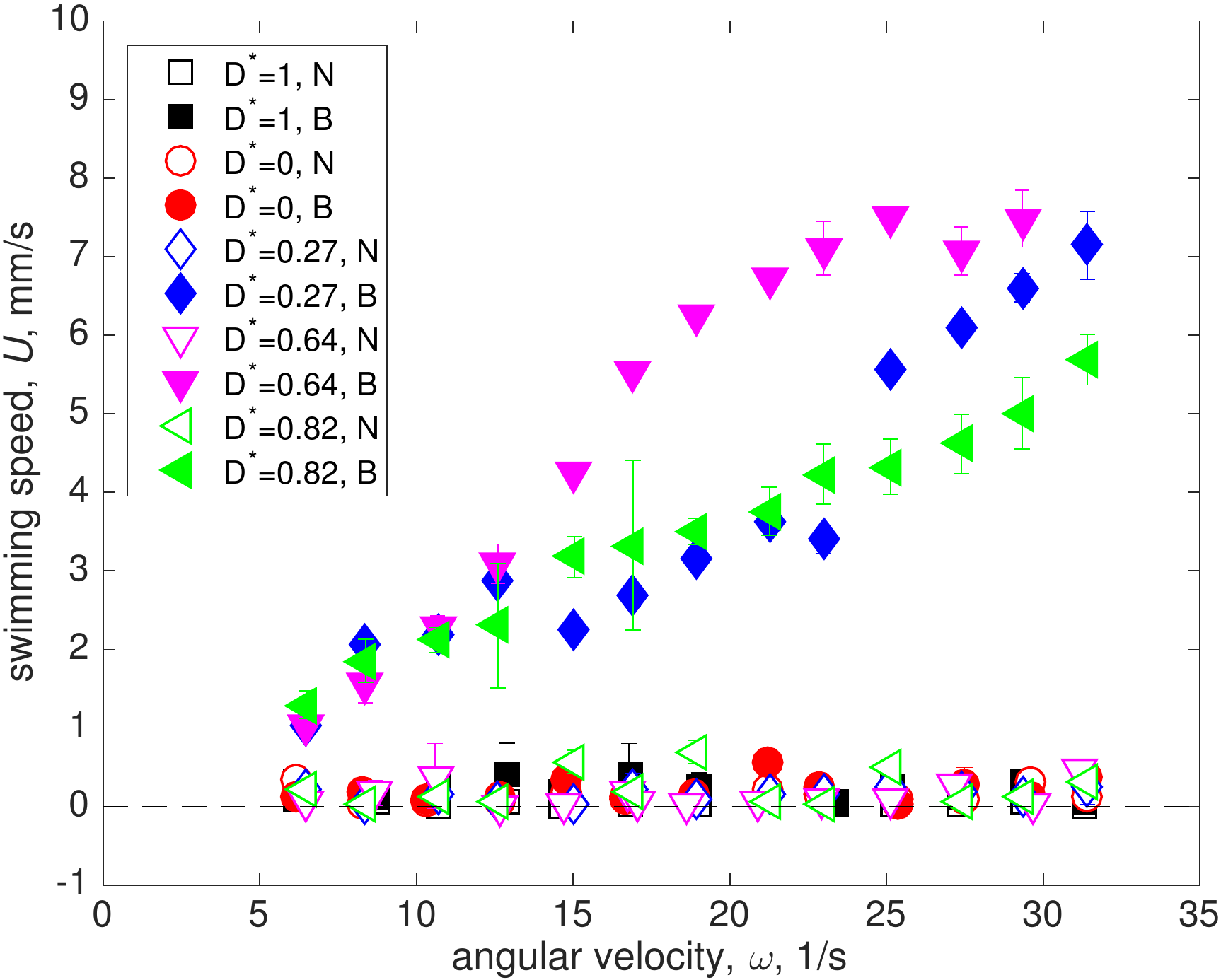}
\caption{Swimming speed of dumbbells,  $U$ (mm/s), as a function angular velocity, $\omega$ (1/s). The different symbols for each experiment correspond to the symbol shapes shown in Table \ref{sphereproperties}. Empty and filled symbols correspond to the data for  the Newtonian fluid (N) and Boger viscoelastic fluid (B), respectively. }
\label{fig:results1}
\end{figure}

\section{Experimental results}
\label{sec:results}

The velocities of the dumbbells were measured in both fluids for the range of rotational speeds indicated above. The results are shown in Fig.~\ref{fig:results1} with empty symbols used for  the data in  the Newtonian fluid (N)
and the solid symbols  for the Boger viscoelastic fluid (B). First, we tested  the limiting case $D^*=0$, corresponding to the limit where the dumbbell is in fact a single sphere. By symmetry a single rotating sphere  is not expected to swim in either fluid, as confirmed experimentally (circular symbols).  Similarly,  the dumbbell with $D^*=1$ (formed by two equal spheres) should not self-propel in any fluid because, again, it is symmetric. This is also confirmed by our measurements (square symbols)

For the other three dumbbells in a snowman configuration ($D^*$= 0.27, 0.64 and 0.82), the swimming speed was confirmed to be zero for the case of the Newtonian fluid, as shown in Fig.~\ref{fig:results1} (empty symbols). The absence of swimming is a result of the time-reversibility of the Stokes equations and the small experimental deviations from  zero in the figure result from  uncertainties in our measurements.

\begin{figure}[t]
\centering
\includegraphics[width=0.6\textwidth]{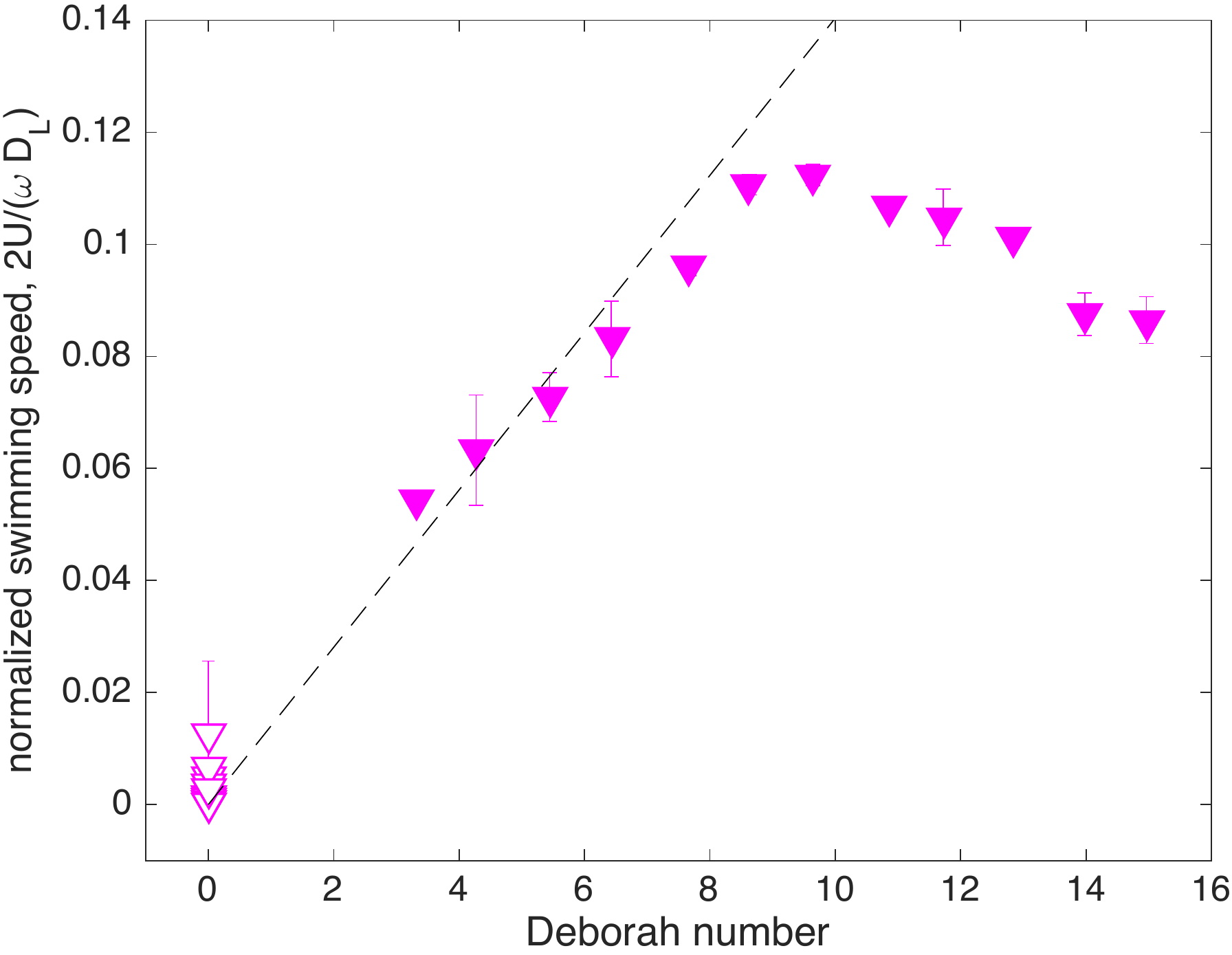}
\caption{Normalized swimming speed,  $2U/(\omega D_L)$, as a function of Deborah number, $\De$. For clarity, we display only the results for the size ratio $D^*= 0.64$. Filled and empty symbols show the normalized swimming speed for the Boger (N) and Newtonian (N) fluids, respectively.The dashed line in the plot shows the trend $U^* \sim \De$, proposed by~\cite{pak12}.}
\label{fig:results2}
\end{figure}

The most important result, which confirms the fundamental idea of this paper, is the non-zero swimming speed measured for case of dumbbells in an asymmetric snowman configuration when immersed in the Boger fluid. In the three cases, the swimming speed is found to monotonically increase with the angular velocity (Fig.~\ref{fig:results1}, filled symbols). The direction of motion is always in the direction from the largest to the smaller sphere, as expected from the difference in viscoelastic force between large and small spheres  explained below.
We next plot in Fig.~\ref{fig:results2}  the normalized swimming speed, $U^*=2U/(\omega D_L)$, as a function of the Deborah number, $\De$ (for clarity we only display the
case for $D^*= 0.64$, the others being similar).
This dimensionless swimming speed increases with Deborah number, reaches a maximum value  after which it  decreases,  in qualitative agreement with  Ref.~\cite{pak12}.  It is notable   that the swimming speeds in both the  experiments and the  model have a maximum  for a certain value of the Deborah number, despite the fact that the model is expected to be valid only for small \De. We also note that for small values of \De \, the swimming speed increases linearly with \De, in agreement with the analytical calculation in Ref.~\cite{pak12}. The dashed line in the plot shows the trend $U^* \sim \De$, showing the good agreement.

\begin{figure}[t]
\centering
\includegraphics[width=0.6\textwidth]{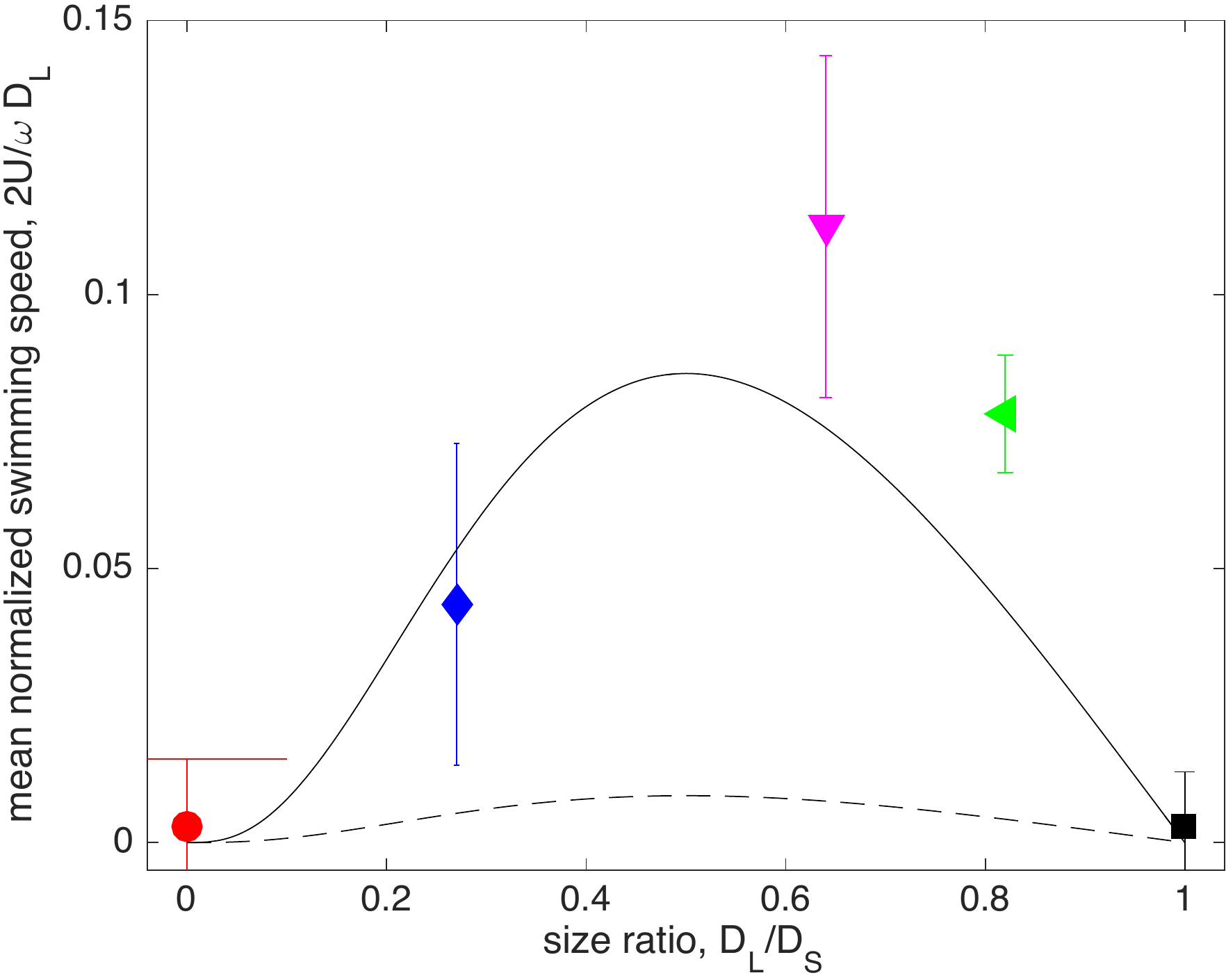}
\caption{Normalized swimming speed, $2U/(\omega D_L)$, as a function of size ratio, $D^*$, for Deborah number $\De=10$. Error bars are $\pm$ one standard deviation and symbols refer to the swimmer geometries in Table~\ref{sphereproperties}. The solid lines shows the analytical model from Pak \etal~\cite{pak12} with no fitting parameters
 (Eq.~\ref{eqn:Ustar} below, which is Eq.~51 in their original study). The dashed line shows the same theoretical prediction but for  a value of $\De=1$  below which  the analytical model would normally be expected to be valid. }
\label{fig:results3}
\end{figure}

As was demonstrated by  Pak \etal~\cite{pak12}, the swimming speed of snowmen is maximized for a specific ratio of sphere sizes ($D^*\approx 0.5$ in that work). The existence of an optimum is a simple consequence of the fact that no swimming is obtained in the limits $D^*= 0$ or 1 and therefore maximum swimming is obtained  at some intermediate value of $D^*$. To demonstrate the existence of an optimal size ratio in our experiments, we plot in Fig.~\ref{fig:results3} our results of the normalized swimming speed as a function of size ratio, $D^*$, for the value
$\De =10$ at which the speed was maximum. Although only five data points are available,
 the existence of an optimal size ratio is clear. Among our experiments, the snowman with $D^*=0.64$ has the largest dimensionless speed.

\section{Modelling}
\label{sec:model}
In their original paper, Pak \etal~\cite{pak12} proposed a physical mechanism to explain the swimming of the snowman based on our understanding of the non-Newtonian flows induced by the rotation of spheres in elastic fluids.
When an isolated sphere is rotated in a viscoelastic liquid, a secondary flow is induced  away from  the sphere along the axis of rotation and towards it in the perpendicular direction \cite{birdvol1}. Since the streamlines are curved, hoop stresses appear which stretch the polymer molecules giving rise to an additional tension along the close stream lines. This tension leads to an inward contraction which results in this secondary flow.  Considering a dumbbell made up of  two widely-separated spheres, the secondary flows created by the rotation each sphere would tend to push the two spheres apart. However,  since the flow strength depends on the sphere size, an unequal sphere pair will result in a net elastic force directed from the large to the small sphere.

This physical argument was verified using the numerical simulations and theory by Pak \etal~in their original article, and the current study allows to confirm these results experimentally. We can  compare our measurements to the mathematical model derived by   Pak \etal~\cite{pak12}. Balancing the net viscoelastic force resulting from the difference between the secondary flows induced by the two spheres (evaluated analytically in the small-$\De$ limit) with the  viscous drag on the dumbbell leads to a prediction for the swimming speed $U$ as
\begin{equation}\label{eqn:Ustar}
\frac{2U}{\omega D_L} =  \De (1-\beta)\frac{2(D^*)^3(1-D^*)}{(1+D^*)^6}
\end{equation}
where $\De$ is de Deborah number (defined in Eqn. \ref{eqn:De}) and $\beta$ is the ratio of solvent to total viscosities (see Table \ref{rheologicalproperties}).

In Fig.~\ref{fig:results3} the prediction from Eq.~(\ref{eqn:Ustar}) is shown, for $\De=10$ and using the value $\beta=0.1149$ from Table \ref{rheologicalproperties}, along with the experimental measurements. Given that the model has   no fitting parameters, and although it is valid only in the limit $\De\ll 1$, the agreement with the experimental measurements is very good.

\section{Conclusions}
\label{sec:conclusion}

Motivated by a recent theoretical proposal, we carried out experiments demonstrating the use of normal stress differences for propulsion. Rigid  dumbbells were rotated by an external magnetic field along their axis of symmetry in a constant-viscosity viscoelastic fluid. When the shape of the dumbbell is asymmetric  (snowman geometry),  normal stress differences  lead to a net propulsion in the direction of the smaller sphere. The use of a simple mathematical  model developed in the limit of widely-separates spheres allows to the rationalize the experimental results by predicting the dependence of the snowman swimming speed on the size ratio between the two spheres.  Although the model is only valid for asymptotically small values of the Deborah number, the good agreement obtained between  experiments and theory indicates that it is able to  capture the relevant physics, namely  propulsion induced by secondary elastic flows. Of course, it  is well known that many mathematical models derived in a specific asymptotic limit continue to be semi-quantitatively correct  outside their domains of validity. The good agreement demonstrates that the model is robust and its use and validity should be explored further.

The results in this paper are valuable because they confirm experimentally an important idea: the viscoelastic nature of a complex  fluid can be exploited in a useful manner. In general, the elastic flow effects are considered a negative feature since their implications are often unclear and detrimental. In this case, however,  elasticity leads to a clear, measurable and well characterized effect: net propulsion resulting from geometrical asymmetry.
Such an effect can give rise to micro-swimmers that can propel themselves in biological fluids (most of which are viscoelastic) for biomedical applications, without complex actuation mechanisms. Other similar attempts with equal-sized-sphere dimers were observed to swim in heterogenous viscoelastic media \cite{Rogowski2018}, but their maneuverability could not be fully controlled. In contrast, for the snowman configuration proposed here, the displacement will always occur in a known direction. Furthermore, as original hypothesised by Pak \etal~\cite{pak12}, the measurement of rheological properties of viscoelastic liquids,  which remains   a technical challenge, could be determined directly from the speed of asymmetric swimmers since there is a direct link between rheology and propulsion characteristics. Finally, the asymmetries of flow configurations could also be used to induce  elastic effects in a controlled fashion, for fundamental fluid mechanic studies. We plan to pursue some of these ideas in the future.

\smallskip

This project has received funding from the European Research Council (ERC) under the European Union's Horizon 2020 research and innovation programme  (grant agreement 682754 to EL). RZ acknowledges the Isaac Newton Institute, Cambridge, for granting a Simons Fellowship; this work was finalised during a stay supported by the fellowship.


\end{document}